\documentclass{article}

\usepackage{arxiv}

\usepackage[utf8]{inputenc} 
\usepackage[T1]{fontenc}    
\usepackage{hyperref}       
\usepackage{url}            
\usepackage{booktabs}       
\usepackage{amsfonts}       
\usepackage{nicefrac}       
\usepackage{microtype}      
\usepackage{lipsum}		
\usepackage{graphicx}
\usepackage{natbib}
\usepackage{doi}
\usepackage{tikz, pgfplots}
\usetikzlibrary{angles}
\usepackage{siunitx}
\usepackage{physics}
\usepackage[ruled]{algorithm2e}

\makeatletter
\newcommand{\mycaption}{%
\ifx \@captype \@undefined \@latex@error {\noexpand \caption outside float}\@ehd \expandafter \@gobble \else \refstepcounter \@captype \expandafter \@firstofone \fi {\@dblarg {\@caption \@captype }}%
}%
\makeatother

\newcommand{\iu}{\mathrm{i}\mkern1mu}

\DeclareSIUnit\angstrom{\text {Å}}

\title{Ghostbuster: a phase retrieval diffraction tomography algorithm for cryo-EM}

\date{} 					

\author{%
\href{https://orcid.org/0000-0001-5160-7628}{\includegraphics[scale=0.06]{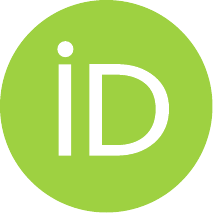}\hspace{1mm}Joel Yeo$^{1,2,3}$} \quad \href{https://orcid.org/0000-0002-1887-7551}{\includegraphics[scale=0.06]{orcid.pdf}\hspace{1mm}Benedikt J. Daurer$^{5,6}$} \quad \href{https://orcid.org/0000-0002-2662-6373}{\includegraphics[scale=0.06]{orcid.pdf}\hspace{1mm}Dari Kimanius$^{7,8}$} \quad \href{https://orcid.org/0000-0001-8684-1158}{\includegraphics[scale=0.06]{orcid.pdf}\hspace{1mm}Deepan Balakrishnan$^{4,5}$}\\
\href{https://orcid.org/0000-0001-5595-9954}{\includegraphics[scale=0.06]{orcid.pdf}\hspace{1mm}\textbf{Tristan Bepler}$^{9}$} \quad 
\href{https://orcid.org/0000-0001-6656-6320}{\includegraphics[scale=0.06]{orcid.pdf}\hspace{1mm}\textbf{Yong Zi Tan}$^{4,5,10,11}$} \quad \href{https://orcid.org/0000-0002-8886-510X}{\includegraphics[scale=0.06]{orcid.pdf}\hspace{1mm}\textbf{N. Duane Loh}$^{1,2,4,5*}$}\\
$^1$NUS Graduate School for Integrative Sciences and Engineering Programme, National University of\\Singapore, 119077 Singapore, Singapore, \\ 
$^2$Department of Physics, National University of Singapore, 117551 Singapore, Singapore,\\
$^3$Institute of Materials Research and Engineering (IMRE), Agency for Science, Technology and
Research\\(A*STAR), 2 Fusionopolis Way, Innovis \#08-03, 138634 Singapore, Singapore,\\
$^4$Department of Biological Sciences, National University of Singapore, 117558 Singapore, Singapore,\\
$^5$Center for Bio-Imaging Sciences, National University of Singapore, 117557 Singapore, Singapore,\\
$^6$Diamond Light Source, Harwell Campus, Didcot, OX11 0DE, UK,\\
$^7$MRC Laboratory of Molecular Biology, Francis Crick Avenue, Cambridge, CB2 0QH, UK,\\
$^8$CZ Imaging Institute, 3400 Bridge Parkway, Redwood City, CA 94065, USA,\\
$^9$Simons Machine Learning Center, New York Structural Biology Center, New York, NY, USA,\\
$^{10}$Disease Intervention Technology Laboratory (DITL), Agency for Science, Technology and Research (A*STAR),\\8A Biomedical Grove, 138648 Singapore, Singapore,\\
$^{11}$Institute of Molecular and Cell Biology (IMCB), Agency for Science, Technology and Research (A*STAR),\\61 Biopolis Drive, Proteos, 138673 Singapore, Singapore\\ \\
$^*$\texttt{Corresponding author: duaneloh@nus.edu.sg}
}

\renewcommand{\shorttitle}{Ghostbuster: a phase retrieval diffraction tomography algorithm for cryo-EM}

\hypersetup{
pdftitle={Ghostbuster: a phase retrieval diffraction tomography algorithm for cryo-EM},
pdfsubject={},
pdfauthor={Joel Yeo, Benedikt J. Daurer, Dari Kimanius, Deepan Balakrishnan, Tristan Bepler, Yong Zi Tan, N. Duane Loh},
}

\begin{document}
\maketitle

\begin{abstract}
	Ewald sphere curvature correction, which extends beyond the projection approximation, stretches the shallow depth of field in cryo-EM reconstructions of thick particles.
    Here we show that even for previously assumed thin particles, reconstruction artifacts which we refer to as ghosts can appear.
    By retrieving the lost phases of the electron exitwaves and accounting for the first Born approximation scattering within the particle, we show that these ghosts can be effectively eliminated.
    Our simulations demonstrate how such ghostbusting can improve reconstructions as compared to existing state-of-the-art software.
    Like ptychographic cryo-EM, our Ghostbuster algorithm uses phase retrieval to improve reconstructions, but unlike the former, we do not need to modify the existing data acquisition pipelines.
\end{abstract}


\section{Introduction}
\label{sec:intro}
Continual advances in cryogenic electron microscopy (cryo-EM) imaging have paved the way toward increasing resolution for single particle reconstruction (SPR) \citep{Kuhlbrandt2014-qp}.
This advance can be attributed to technological developments in instrumentation that have continued to push the resolution limits in cryo-EM \citep{Nakane2020-lo, yip2020atomic}.
However, the depth of field (DOF) of the microscope decreases with increasing resolution \citep{Du2020-zs}, which narrows cryo-EM's DOF as it trends toward higher resolutions \citep{DeRosier2000-pd}.

In this scenario, the projection approximation (PA) that is commonly used in cryo-EM reconstruction must give way to approaches that account for the diffraction physics within the sample \citep{Vulovic2014-nd}.
Such approaches are sometimes referred to as diffraction tomography \citep{Muller2015-nl}, which computationally extends shallow DOFs.
There are two common approaches, which are especially important for thick samples.
The first approach is anchored in the multislice formalism in real space \citep{Cowley1957-jt}.
Specifically, a thick sample is segmented into multiple 2D slices, where an incoming electron wave serially transmits and propagates through these slices.
If we can recover this complex-valued electron exitwave, then we can extend the narrow DOF by computationally backpropagating the wave through the sample (similar to what is done in \citep{Muller2015-nl, Ren2020-xg, Chen2021-wl, Gureyev2022-gt}).
The second approach is described by the curvature of the Ewald sphere in the Fourier domain \citep{DeRosier2000-pd}.
Ewald sphere curvature correction (ESCC) algorithms have been developed to stretch this shallow DOF \citep{Wolf2006-dr, Russo2018-gd} with some experimental success \citep{Tan2018-uv, Nakane2020-lo, Tegunov2021-cs}.
However, ESCC requires a disambiguation procedure \citep{Wolf2006-dr, DeRosier2000-pd} that, as we show below in Sec.~\ref{sec:equivalence}, can be resolved if the complex-valued exitwaves are known.

However, these complex-valued exitwaves are not directly measurable.
Instead, they propagate through the microscope optics and accumulate aberrations, resulting in a complex-valued interference pattern formed at the detector known as an in-line hologram \citep{Wade1992-bs}.
Since a detector only captures the intensities of these holograms, their phases are lost upon measurement as bright field electron micrographs.
Therefore, recovering the complex-valued exitwaves (crucial to the two approaches in the previous paragraph) requires both the retrieval of their holograms' lost phases and the removal of microscopy aberrations.
This `missing phases problem' is common in many fields of imaging, including x-ray coherent diffractive imaging, electron microscopy, and astronomy \citep{Shechtman2015-li}.

Over the last few decades, many have shown that in certain conditions these missing phases can be computationally retrieved \citep{Gerchberg1972-ur, Fienup1982-xf, Elser2003-rf, Candes2013-px, Candes2015-vm}.
Cryo-EM has circumvented computational phase retrieval with a first-order linearization of the image formation physics \citep{Wade1992-bs}, which is succinctly captured by a contrast transfer function (CTF).
This linearized model has sufficed to achieve near-atomic resolution cryo-EM reconstructions via CTF correction algorithms \citep{Penczek2010-iw, singer2020computational, Nakane2020-lo, yip2020atomic}.
However, the phases of the complex-valued exitwaves are unattainable in this first-order linearized model.
Without these phases, we cannot stretch the DOF of cryo-EM reconstructions beyond the current state of the art \citep{Heymann2023-jp}.
Incidentally, computational phase retrieval, through ptychography \citep{Zhou2020-uz, pelz2017low, Pei2023-ru}, has been increasingly recognized as a possible route to improve imaging contrast for cryo-EM.
However, in cryo-EM these ptychographically-retrieved phases have not been exploited to stretch this DOF.

There have been recent efforts to computationally retrieve (whether explicitly or implicitly) the complex-valued exitwaves to enable diffraction tomography.
By accounting for dynamic scattering, Ren \emph{et al.} \citep{Ren2020-xg} demonstrated improved 3D reconstructions at much lower electron doses for phase-contrast atomic electron tomography.
However, a more computationally efficient first Born approximation of diffraction tomography (i.e. kinematic scattering) can already reduce artifacts (similar to the ghosts shown below in Figs.~\ref{fig:sphericalparticle},~\ref{fig:compare_size},~\ref{fig:compare_ice} and~\ref{fig:recon}) observed in computed tomography, even for small particles \citep{Gureyev2020-wo, Gureyev2021-eq}. 
Similar demonstrations were extended to simulated cryo-EM \citep{Gureyev2021-eq, Chen2021-wl}.
Nevertheless, an end-to-end cryo-EM reconstruction pipeline using this kinematic scattering approximation \citep{Chen2021-wl} has a large memory overhead.
This overhead can be circumvented by first recovering the 2D complex-valued exitwaves of individual particles, followed by 3D reconstruction.
However, this two-step approach requires first retrieving the phases of individual cryo-EM particles.
Earlier phasing attempts by Gureyev \emph{et al.} needed at least two defocus pairs per orientation \citep{Gureyev2020-wo, Gureyev2021-eq}, which is impractical for cryo-EM.
This defocus pair requirement was subsequently removed in their later work \citep{Gureyev2022-gt} but yielded visibly poorer results.

Here, we address the aforementioned challenges in the algorithm that we have developed here: Ghostbuster.
Ghostbuster maintains a low memory footprint by taking a two-step approach of phase retrieval followed by 3D reconstruction based on minibatch stochastic gradient descent (SGD) for diffraction tomography.
Unlike Gureyev \emph{et al.}, we simultaneously retrieve the phases for all particles by constraining their recovered exitwaves to concur as a single 3D particle.
We show that this phasing approach sets the upper bound of state-of-the-art CTF correction methods \citep{Penczek2010-iw, singer2020computational, Nakane2020-lo, yip2020atomic}.
3D reconstruction is then performed by minimizing the error between the first Born approximation of the multislice formalism and the phase-retrieved exitwaves via minibatch SGD.
This reduces the large memory overhead by only using a small subset of the full dataset per iteration, whilst also improving the reconstruction by avoiding local minima.
Using a multislice simulated particle stack, we demonstrate improved 3D reconstruction with Ghostbuster as compared to RELION \citep{Kimanius2021-gd}.
In addition, we model the impact of these artifactual ghosts on resolving power, even for small particles.
This is, in turn, used to empirically define the DOF of a coherent imaging system.
The removal of these ghostly artifacts, which we call ghostbusting, moves us toward thicker samples, and automatic masking.

\section{Theory: resolution considerations in diffraction tomography}
\label{sec:theory}

\subsection{Depth of field when coherently imaging multiple atoms}
\label{sec:dof}

The PA assumes that an imaging system has infinite DOF.
In reality, the effects of extended defocus within a thick, 3D particle lead to reconstructions with a shallow DOF.
This shallowness is due to the accumulation of defocus as an electron wave propagates through the particle, which we will later show (as others have \citep{Vulovic2014-nd, DeRosier2000-pd}) to be equivalent to the curvature of the Ewald sphere.
Reconstructing an object to the highest fidelity requires accounting for this range of defocus within the object. 
Without doing so, artifacts, which we call ghosts, will manifest even for small particles \citep{Gureyev2020-wo, Gureyev2021-eq}.

\begin{figure}[htbp]
    \centering
    \includegraphics[width=\textwidth]{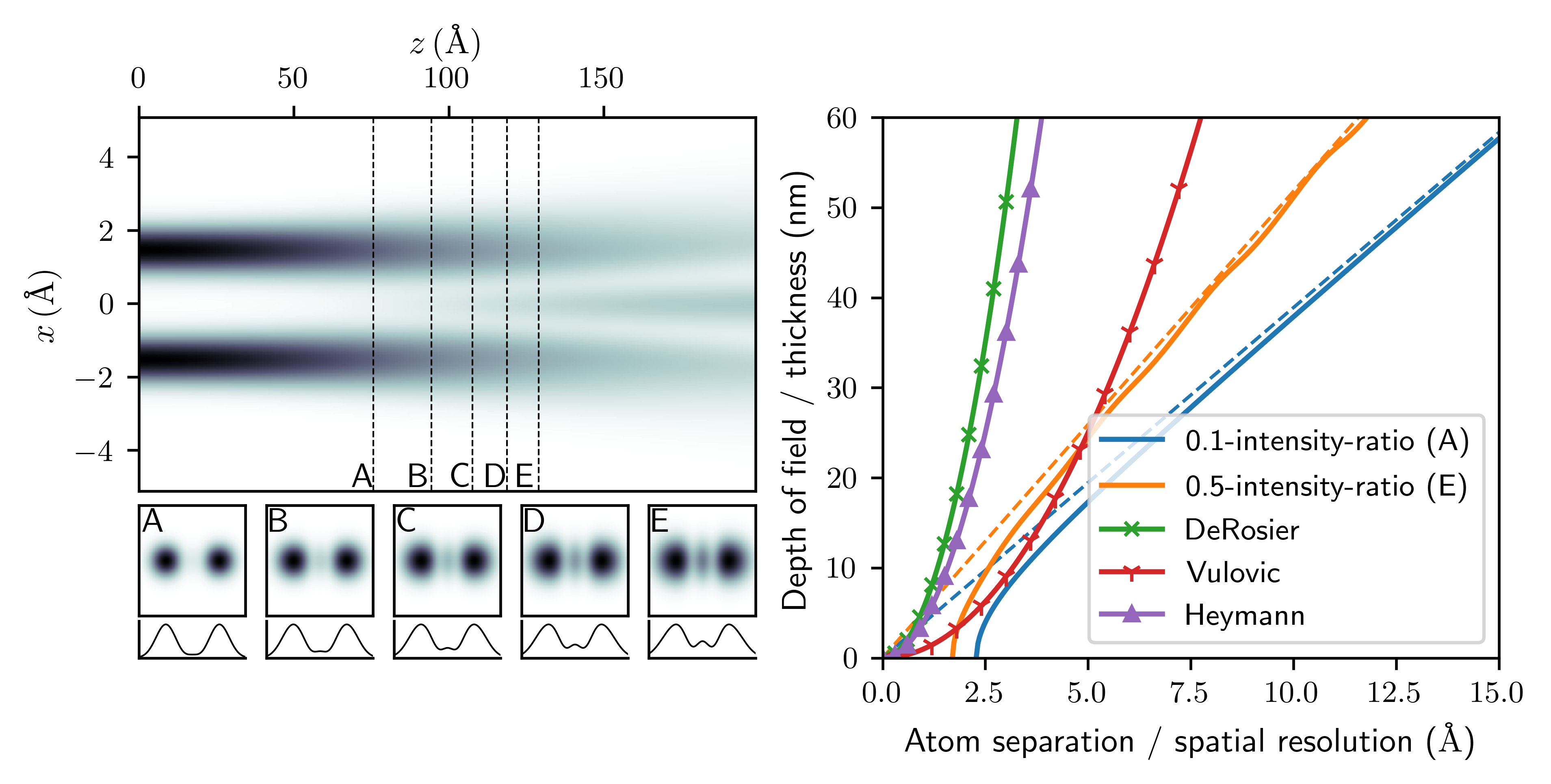}
    \caption{Coherent criteria for resolving two atoms without ghosts. 
    (Left) Propagation of an electron wave arising from two Gaussians with a full-width at half maximum (FWHM) of \SI{1.4}{\angstrom} (standard deviation of \SI{0.59}{\angstrom}) each, separated by a distance of \SI{3}{\angstrom}. 
    The lower smaller panels depicts the cross-section views of the electron wave intensity at propagated distances of (A) \SI{75.7}{\angstrom}, (B) \SI{94.3}{\angstrom}, (C) \SI{107.6}{\angstrom}, (D) \SI{118.8}{\angstrom}, and (E) \SI{128.9}{\angstrom}. 
    These distances correspond to the location where the ratio of the ghost to the actual atoms' peak intensities are equal to 0.1/0.2/0.3/0.4/0.5. 
    (Right) The DOF (or particle thickness) as a function of atom separation (or resolution). 
    The asymptotes of our ghost-based resolution criteria (dashed lines) depict a linear trend as opposed to the quadratic relationship (solid lines) suggested by previous works \citep{DeRosier2000-pd, Vulovic2014-nd, Heymann2023-jp}.
    Equations of these asymptotes can be found in the Supplementary Information.}
    \label{fig:resolution}
\end{figure}

Previous works estimated the DOF from the broadening of exitwaves from low-resolution features \citep{DeRosier2000-pd, Vulovic2014-nd, Heymann2023-jp}.
These estimates should be revisited as cryo-EM approaches atomic resolution.
At such high resolutions, we have to consider the interference of the scattered electron wave between multiple atoms.
This careful examination is illustrated in Fig.~\ref{fig:resolution}: the free-space propagation of a \SI{200}{\keV} electron plane wave passing through two actual atoms (modeled as 2D Gaussians at the $z=0$ plane with a size of \SI{0.7}{\angstrom}) produces a central ghost atom due to interference.
We empirically define the DOF as the propagation distance for which the central ghost atom's peak intensity becomes noticeable with respect to the actual atoms' peak intensities (i.e., their intensity-ratios exceed 0.1; details in Supplementary Information).

For a range of atomic separations, we plot this DOF for various noticeability criteria in Fig.~\ref{fig:resolution}.
Putting this into context, to resolve a pair of atoms within a particle to \SI{3}{\angstrom} resolution without significant ghosts requires a DOF that is less than $\sim$\SI{100}{\angstrom}.
If the DOF is smaller than the particle of interest, then PA-based reconstructions will produce a 3D estimate plagued with ghost atoms (comparable to Figure 5 in \citep{Gureyev2021-eq}).
Interestingly, this DOF becomes even shallower when quantum mechanical effects of scattering \citep{Forbes2010-gl} are included (see Supplementary Fig. 1).

Fig.~\ref{fig:resolution} shows that previous estimates of resolution, $r$, vs DOF (or thickness, $t$)  scales like $t\propto r^2$ \citep{DeRosier2000-pd, Vulovic2014-nd, Heymann2023-jp}, whereas our ghost-based resolution criterion scales like $t \propto r$.
This difference is largely because previous estimates do not consider artifacts resulting from interference between multiple atoms, which results in a larger DOF estimate.

\begin{figure}[htbp]
    \centering
    \includegraphics[width=0.5\textwidth]{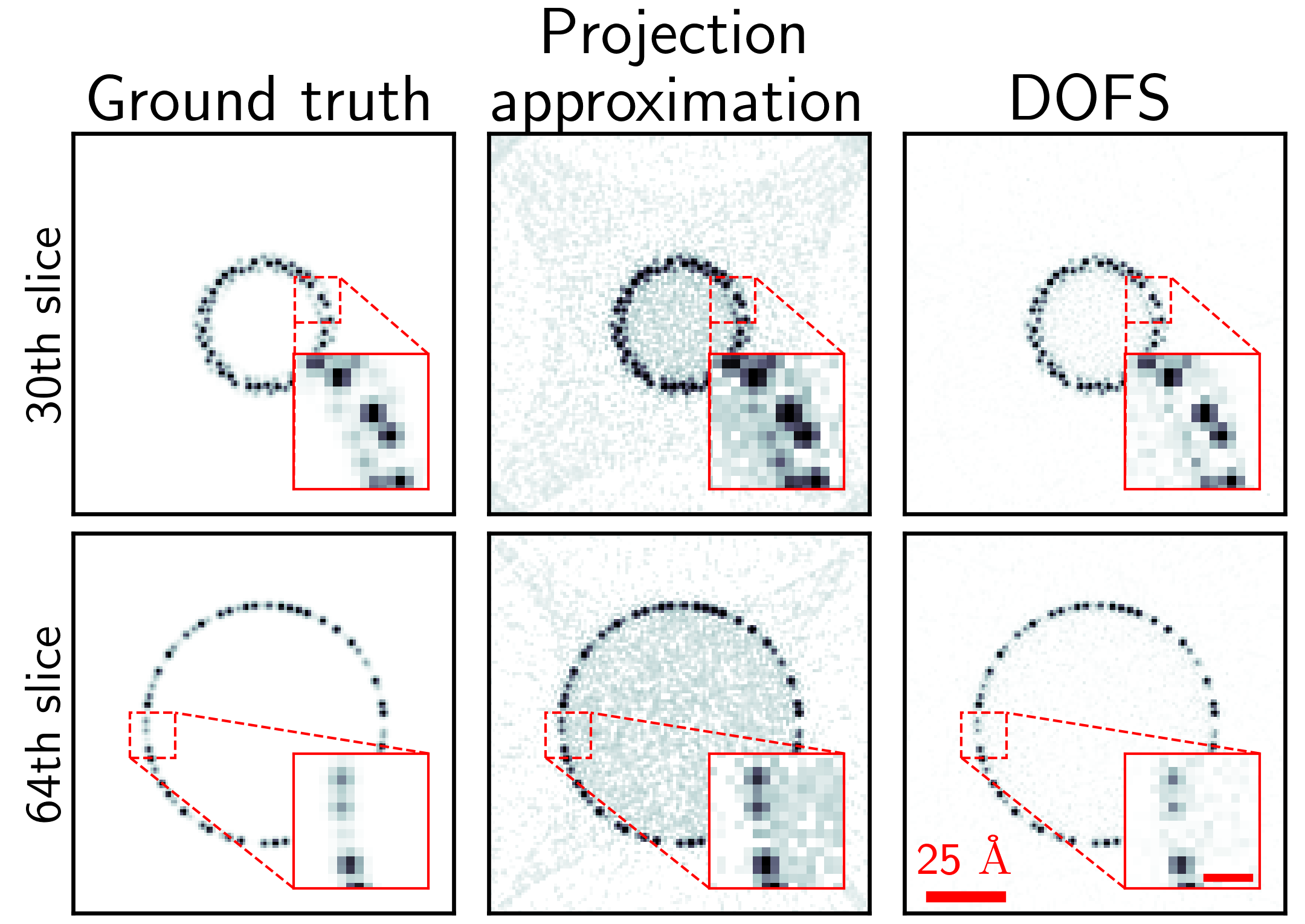}
    \caption{2D slices of a 3D reconstruction for a spherical particle. 
    The scale bar within the inset corresponds to \SI{5}{\angstrom}.
    Without accounting for diffraction within the particle, the PA reconstruction results in many ghost atoms, including the central region of the particle which should not have atoms.
    These ghosts are `busted' when we use our depth of field stretching (DOFS) algorithm (see Sec.~\ref{sec:DOFS}) to reconstruct instead.}
    \label{fig:sphericalparticle}
\end{figure}

To illustrate these differences, consider the spherical phantom of diameter $\sim$\SI{100}{\angstrom} depicted in Fig.~\ref{fig:sphericalparticle}.
According to DeRosier's rule-of-thumb, which is used in RELION \citep{10.7554/eLife.42166} and cryoSPARC \citep{Punjani2017-oi}, this phantom should be fully in focus (i.e., PA is valid) at $\sim$\SI{1.25}{\angstrom} resolution.
However, Fig.~\ref{fig:sphericalparticle} shows how assuming the PA for this phantom results in ghosts because the diffraction physics within the particle was unaccounted for.
This suggests that we should revisit the resolution criteria for particles that are currently considered thin.

\subsection{Propagation and reconstruction within a thick particle}

\subsubsection{Multislice description of beam propagation.}

The multislice formalism is an established numerical model of the diffraction physics within the particle \citep{Cowley1957-jt, Kirkland2010-kj}.
For completeness, we describe how the multislice model simulates an exitwave produced when an incident electron plane wave propagates through a particle.
Let this particle's 3D complex-valued scattering potential $v$, be represented by $M$ equally spaced 2D slices (i.e., $\qty{v_1,\ldots, v_m, \ldots, v_M}$), along the beam propagation axis, $z$.
The exitwave is calculated via the recurrence relation:
\begin{align}\label{eq:multislice}
    \psi_{m+1}\qty(\vb{x}) = \qty[\psi_m\qty(\vb{x}) \otimes p\qty(\vb{x})] \exp[\iu \sigma_e v_{m+1}\qty(\vb{x})],
\end{align}
where $\psi_m$ is the electron wave just after the $m$th slice; $\qty(\vb{x})=\qty(x,y)$ denotes the spatial coordinates perpendicular to the beam propagation axis; $\sigma_e$ is the interaction parameter; $\otimes$ is the convolution operator; and $p$ is the Fresnel propagator whose Fourier transform is defined as
\begin{align}\label{eq:fresnel}
    P\qty(\vb{k}) = \exp(\iu \pi \Delta z \lambda k^2),
\end{align}
where $\qty(\vb{k})=\qty(k_x,k_y)$ denotes the 2D spatial frequency coordinates, $\Delta z$ is the distance between each slice, $\lambda$ is the wavelength of the electron beam, and $k=\qty|\vb{k}|$.
Eq.~\eqref{eq:multislice} describes the repeated process of the electron wave undergoing transmission and free-space propagation until it reaches the exitplane at $m=M$.

The first Born approximation of the multislice formalism allows us to further simplify Eq.~\eqref{eq:multislice} to \citep{Balakrishnan2023-au}
\begin{align}\label{eq:linear_multislice}
    \psi_\mathrm{exit}\qty(\vb{x}) \approx 1 + \iu \sigma_e \sum_{m=1}^M v_m\qty(\vb{x}) \otimes p_{m}\qty(\vb{x}),
\end{align}
where $\psi_\mathrm{exit}$ is the exitwave, and the Fourier transform of $p_m$ is instead
\begin{align}
    P_m\qty(\vb{k}) = \exp[\iu \pi \qty(M-m)\Delta z  \lambda k^2].
\end{align}
In this approximation, a plane wave is assumed to be incident on each independent slice.
The weak-phase approximation is also used to linearize the transmission function for each slice: $\exp\qty(\iu\sigma_e v_m)\approx 1 + \iu \sigma_e v_m$.
The resultant scattered wave from each slice propagates directly to the exitplane without passing through the other slices.
The contributions from each slice are then coherently summed at the exitplane.
Therefore, Eq.~\eqref{eq:linear_multislice} ignores multiple scattering but retains the relative defocus due to the thickness of the particle.
This first Born approximation of the multislice formalism is also known as the kinematic scattering.

\subsubsection{Kinematic scattering and Ewald sphere curvature.}
\label{sec:equivalence}

Here we show how the spread of defocus in Eq.~\eqref{eq:linear_multislice} can be treated as a curvature of the Ewald sphere.
Consider the 3D scattering potential, $v$, which Fourier transforms to
\begin{align}\label{eq:3dfourier}
    V\qty(k_x, k_y, k_z) &= \iiint v\qty(x,y,z)e^{-i2\pi\qty(k_xx + k_yy + k_zz)}\dd{x}\dd{y}\dd{z} \nonumber \\
    &= \int V\qty(\vb{k},z) e^{-i2\pi k_zz} \dd{z},
\end{align}
where $V\qty(\vb{k},z) = \iint v\qty(\vb{x},z) e^{-i2\pi\qty(k_xx + k_yy)} \dd{x}\dd{y}$.
Comparatively, this same $V\qty(\vb{k},z)$ term also occurs in the Fourier transform of Eq.~\eqref{eq:linear_multislice}:
\begin{align}\label{eq:linearmultislicefourier}
    \Psi\qty(\vb{k}) &= \delta\qty(\vb{k}) + \iu \sigma_e\sum_{m=1}^M V_m\qty(\vb{k})e^{i\qty(M-m)\Delta z \pi \lambda k^2} \nonumber \\
    & \approx \delta\qty(\vb{k}) + \iu \sigma_e e^{\iu M \Delta z \pi \lambda k^2}\int_0^{M \Delta z} V\qty(\vb{k},z)e^{-i z \pi \lambda k^2} \dd{z},
\end{align}
where the approximation refers to the replacement of the summation with an integral in the limit of small $\Delta z$.
The prefactor in Eq.~\eqref{eq:linearmultislicefourier} represents an additional defocus due to our choice of phase origin (i.e. at the $m=1$ slice).
Ignoring this phase origin and the overall scaling factor, we observe an equivalence between Eq.~\eqref{eq:3dfourier} and~\eqref{eq:linearmultislicefourier} if $k_z=\frac{\lambda k^2}{2}$ and $k>0$:
\begin{align}\label{eq:equiv}
    \Psi\qty(\vb{k}) &\approx V\qty(\vb{k}, k_z=\frac{\lambda k^2}{2})
\end{align}

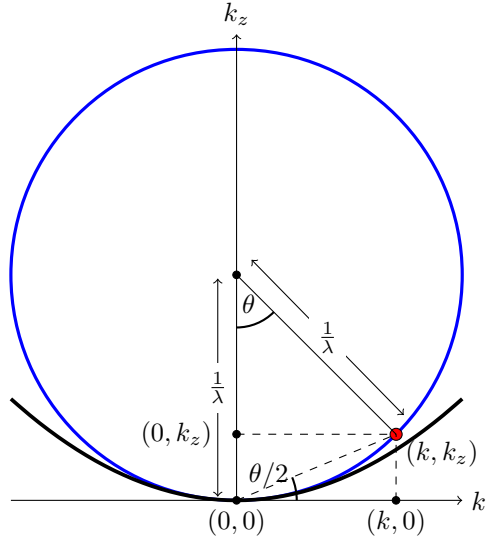
\begin{figure}[htbp]
    \centering
    \begin{tikzpicture}
        \draw[very thick, blue] (0,0) circle (3);

        \draw[<->] (-40:2.9) -- node[midway,fill=white] {$\frac{1}{\lambda}$} (0.2,0.2);
        \draw[<->] (-0.25,-0.05) -- node[midway,fill=white] {$\frac{1}{\lambda}$} (-0.25,-2.95);

        \draw[->] (0,-3) -- (0,3.2) node[above] {$k_z$};
        \draw[->] (-3,-3) -- (3,-3) node[right] {$k$};

        \draw[fill=black] (0,0) circle (0.05);
        \draw[fill=black] (0,-3) circle (0.05) node[below] {$(0,0)$};
        \draw[fill=red] (-45:3) circle (0.08) node[right, yshift=-0.25cm] {$(k,k_z)$};
        \draw[fill=black] (2.12,-3) circle (0.05) node[below] {$(k,0)$};
        \draw[fill=black] (0,-2.12) circle (0.05) node[left, xshift=-0.2cm] {$(0,k_z)$};

        \draw[](0,0) -- (-45:3);

        \draw[dashed] (2.12,-3) -- (-45:3);
        \draw[dashed] (0,-2.12) -- (-45:3);
        \draw[dashed](0,-3) -- (-45:3);

        \draw (-45:3) coordinate (A)  (0,0) coordinate (B)
              (0,-3) coordinate (C)
               pic [draw,black,thick,angle radius=0.7cm, pic text=$\theta$] {angle = C--B--A};
        \draw (-45:3) coordinate (A)  (0,-3) coordinate (B)
              (3,-3) coordinate (C)
               pic [draw,black,thick,angle radius=0.8cm, angle eccentricity=1.5, pic text=$\theta/2$, pic text options={shift={(-0.75, 0.1)}}] {angle = C--B--A};

       \draw[black, line width = 0.50mm]   plot[smooth,domain=-3:3] (\x, {0.15*\x*\x-3});
    \end{tikzpicture}
    \caption{
    Geometrical equivalence of the Ewald sphere curvature (blue) and the first Born approximation of the multislice formalism (black).
    The black curve represents the complex-valued exitwave, $\psi_\mathrm{exit}$, that is mapped onto a paraboloid in Fourier space (Eq.~\eqref{eq:equiv}).
    At a small angle, $\theta\approx \lambda k$, the surface of the paraboloid approaches that of the Ewald sphere since $k_z = \tan \frac{\theta}{2} \approx \frac{k\theta}{2} = \frac{\lambda k^2}{2}$.
    For example, for a \SI{200}{\keV} electron beam ($\lambda\approx$ \SI{0.025}{\angstrom}) and $k=$\SI{1}{\angstrom^{-1}}, the angle is $\theta\approx$ \SI{0.025}{\radian}.
    }
    \label{fig:ewald}
\end{figure}

Eq.~\eqref{eq:equiv} shows that if the complex-valued exitwave, $\psi_\mathrm{exit}$, is known, then its Fourier transform, $\Psi$ (on the left-hand side of Eq.~\eqref{eq:equiv}), can be simply inserted as a curved Ewald surface in the object's Fourier volume (Fig.~\ref{fig:ewald}).
However, since the phases of $\psi_\mathrm{exit}$ are unknown, more sophisticated schemes have been developed to correct for the curvature of the Ewald sphere \citep{Wolf2006-dr, Russo2018-gd}.

\subsubsection{Limitations of the projection approximation.}
In contrast, the PA neglects the extended range of defocus within the particle.
Instead, these slices are projected onto a single $z$-slice in the center of the particle's scattering potential \citep{Balakrishnan2023-au}.
Therefore, the exitwave from this central $z$-slice is approximated to be
\begin{align}\label{eq:pa}
    \psi^\mathrm{PA}_\mathrm{exit}\qty(\vb{x}) \approx 1 + \iu \sigma_e \qty[\sum_{m=1}^M v_m\qty(\vb{x})] \otimes p_z\qty(\vb{x}) = 1 + \iu\sigma_e v_z\qty(\vb{x})\otimes p_z\qty(\vb{x}),
\end{align}
where $v_z = \sum_{m=1}^M v_m\qty(\vb{x})$ is the projected potential, and $p_z$ propagates this slice to the exitplane. 
For a spherical particle, the Fourier transform of $p_z$ can be shown to be \citep{Balakrishnan2023-au}
\begin{align}
    P_z\qty(\vb{k}) = \exp(\iu \pi \frac{M}{2}\Delta z \lambda k^2).
\end{align}

Recall that the criteria defined in Section~\ref{sec:dof} relates a reconstructed particle's DOF with its resolution. 
Consequently, as we increase the resolution, the depth (i.e. thickness) for which Eq.~\eqref{eq:pa} is a good approximation for Eq.~\eqref{eq:linear_multislice} shrinks.
Nevertheless, at a particular resolution, we can stretch the reconstruction's corresponding DOF by accounting for the diffraction physics within the particle (see Section \ref{sec:DOFS}).
The effects of this stretching are evident in Fig.~\ref{fig:sphericalparticle} for a \SI{128}{\angstrom} sized particle.
Ghost atoms appear in the 2D slices of the reconstruction when the PA is assumed even for such a small particle (similar to reports in \citep{Gureyev2020-wo}).
 
\subsubsection{Depth of field stretching (DOFS).}
\label{sec:DOFS}

Stretching the DOF requires solving an inverse problem: reconstructing the $v_m$ slices from a known $\psi_\mathrm{exit}$ based on Eq.~\eqref{eq:linear_multislice}.
This can be rearranged into a minimization problem:
\begin{align}\label{eq:loss}
    \min_{v_m} \norm{\psi_\mathrm{exit} - 1 - \iu\sum_{m=1}^M  v_m\qty(\vb{x}) \otimes p_{m}\qty(\vb{x})}^2_2 \equiv \min_{v_m} E\qty(v_m),
\end{align}
where $E$ is the squared L2-norm loss.
Taking the derivative of this loss function with respect to the $m$th slice, we have
\begin{align}\label{eq:gradient}
    \pdv{E}{v_m} = -2\iu \qty[\psi_\mathrm{exit} - 1 - \iu\sum_{m=1}^M  v_m\qty(\vb{x}) \otimes p_{m}\qty(\vb{x})]\otimes^{-1} p_{m}\qty(\vb{x}),
\end{align}
where $\otimes^{-1}$ is the deconvolution operator.
The gradient in Eq.~\eqref{eq:gradient} is essentially the backpropagation of the difference between the known and estimated exitwave to the respective plane of the $m$th slice. 
This allows us to use iterative gradient descent to reconstruct the $v_m$ slices where the $k$th update is
\begin{align}\label{eq:gradientdescent}
    v_m^{k+1} = v_m^{k} - \mu \pdv{E}{v_m},
\end{align}
where $\mu$ is a user-defined stepsize.
We also enforce a positivity constraint on $v_m^{k}$ at every iteration to improve convergence.

\begin{algorithm}[hbt!]
    \mycaption{Depth-of-field stretching (DOFS)}\label{alg:dofs}
    \KwIn{Known complex-valued 2D exitwaves $\psi_{\mathrm{exit},n}$, rotation matrices $R_n$, stepsize $\mu$, maximum iteration $N_k$, batchsize $N_b$, number of exitwaves $N$}
    \KwOut{Reconstructed 3D map $v$}
    
    {\bf Initialization:} Initial volume estimate: $v^0=0$, initial batch index: $b = 1$
    
    \For{$k = 1 \text{ to } N_k$}{
        
        Reset current gradient: $g = 0$;
        
        \For{$n = b \text{ to } \qty(b + N_b)$}{
    
            Rotate volume:         
            $v^\prime_n = R_{n}\qty{v}$.
            
            Compute first Born approximation exitwave using Eq.~\eqref{eq:linear_multislice}: 
            $\psi_{\mathrm{exit},n}^\prime =  1 + \iu\sum_{m=1}^M  \qty(v^\prime_n)_m \otimes p_{m}.$
    
            Calculate loss against known exitwave using Eq.~\eqref{eq:loss}:
            $E_n = \qty(\psi_{\mathrm{exit},n} - \psi_{\mathrm{exit},n}^\prime)^2$.
    
            Compute the gradients for each slice (Eq.~\eqref{eq:gradient}) and store them as a 3D gradient array:
            $g^{\prime}_n = \qty[\pdv{E}{v_1}, \pdv{E}{v_2}, \ldots, \pdv{E}{v_M}]$.
    
            Backrotate the 3D gradient and add it to the current gradient:
            $g = g + R_{n}^{-1}\qty{g_n^{\prime}}$.
        }
        Average the current gradient:
        $g = \frac{g}{N_b}$.
    
        Update volume:
        $v^{k+1} = v^k - \mu g$.
    
        Positivity constraint:
        $v^{k+1}\qty(x,y,z) = 
        \begin{cases}
            v^{k+1}\qty(x,y,z) &\text{ if } v^{k+1}\qty(x,y,z) \geq 0, \\
            0 &\text{ otherwise }.
        \end{cases}$
        
        Update next batch starting index:
        $b = b + N_b$.
        
    }
\end{algorithm}

However, using just a single 2D complex-valued exitwave to reconstruct a 3D particle is an ill-posed optimization problem.
We can overcome this if multiple exitwaves, which exits from different orientations of the same particle, are known.
This allows us to implement the above approach as a minibatch stochastic gradient descent algorithm, named `depth of field stretching' (DOFS), which is outlined in Algorithm~\ref{alg:dofs}.
Therefore, assuming that we fully know the complex-valued exitwaves and the orientations of the associated particles, DOFS stretches the reconstruction's DOF by accounting for the diffraction physics within the particle.
This is evident in the recovery of the original 3D potential of the spherical phantom in Fig.~\ref{fig:sphericalparticle} to a higher fidelity compared to traditional tomography.



\subsection{Recovering the exitwaves for DOFS}

\subsubsection{Phases are lost upon image formation in cryo-EM.}

\begin{figure}[htbp]
    \centering
    \includegraphics[width=\textwidth]{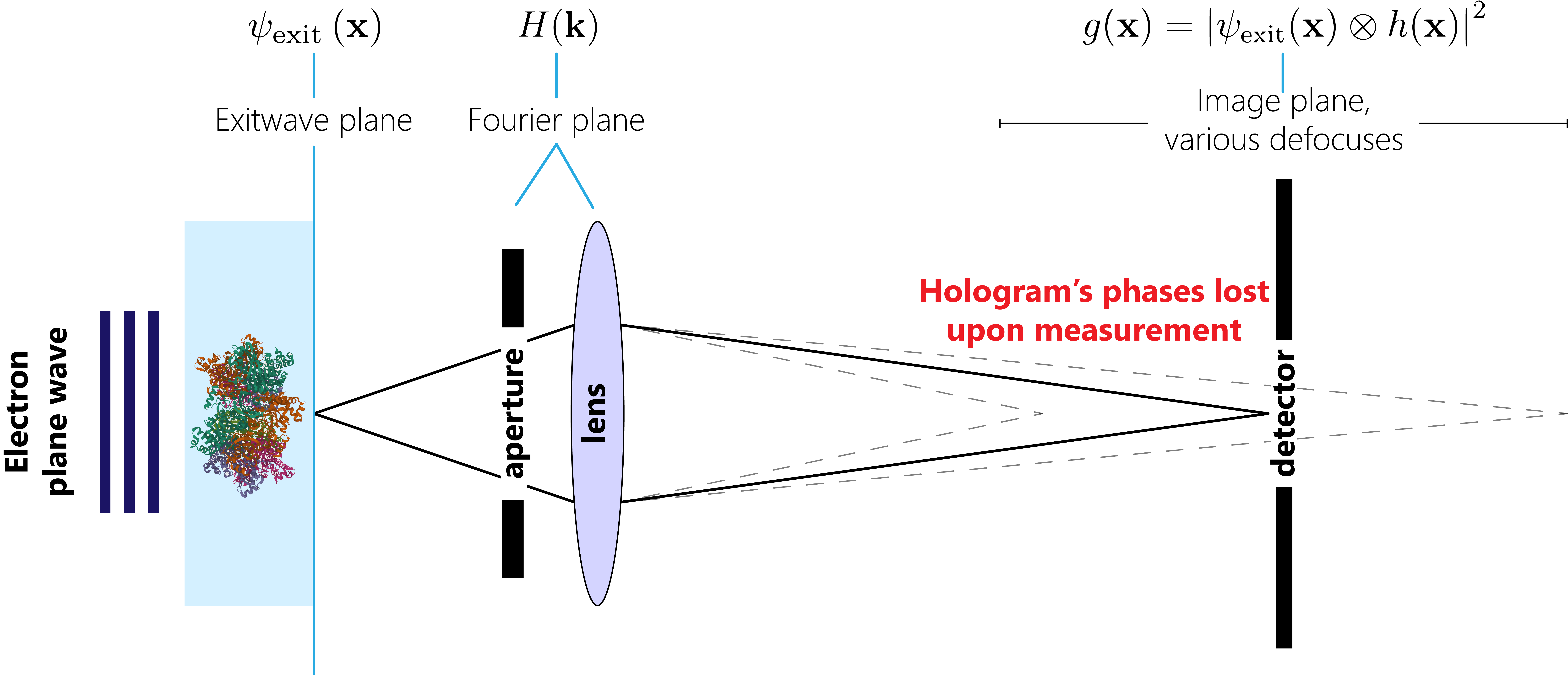}
    \caption{
    Simplified model for cryo-EM imaging.
    An electron plane wave propagates through a 3D particle and emerges as a 2D complex-valued exitwave, $\psi_\mathrm{exit}$, at the exitplane.
    The aberrations accumulated as the exitwave propagates through the microscope optics are described by the transfer function, $H$.
    A complex-valued hologram is formed at the image plane.
    However, the phases of this complex-valued hologram is lost as a detector can only measure its intensity, $g$.}
    \label{fig:cryo-EM}
\end{figure}

The DOFS in the previous section is not immediately possible because we do not measure the complex-valued exitwaves in cryo-EM.
To see this, consider how these exitwaves propagate through a microscope and are finally focused onto a detector as in-line holograms (see schematic in Fig.~\ref{fig:cryo-EM}).
The image that a detector captures, $g$, is only the intensities (and not the phases) of these complex-valued holograms, $\psi_\mathrm{det}$:
\begin{align}\label{eq:g}
    g\qty(\vb{x}) &= \qty|\psi_\mathrm{exit}\qty(\vb{x}) \otimes h\qty(\vb{x})|^2 \equiv \qty|\psi_\mathrm{det}\qty(\vb{x})|^2,
\end{align}
where $h$ and $\psi_\mathrm{det}$ are the complex-valued point-spread function of the microscope and hologram respectively.
A common model for $h$ in cryo-EM is described by its Fourier transform, $H$:
\begin{align}\label{eq:H}
    H\qty(\vb{k}) = \exp[\iu \pi\qty(-\Delta f \lambda k^2 + \frac{1}{2}C_s \lambda^3 k^4)],
\end{align}
where $\Delta f$ is the defocus defined with respect to the exitplane, and $C_s$ is the spherical aberration parameter.
Note that for DOFS to succeed, we must be careful when defining the origin of $\Delta f$.

It is instructive to compare with the effective defocus, $\Delta f_\mathrm{eff}$, typically inferred from cryo-EM images where the PA is used.
According to Eq.~\eqref{eq:pa}, this $\Delta f_\mathrm{eff}$ tends towards the center $z$-slice of the particle, and is different from $\Delta f$ in Eq.~\eqref{eq:H} above in that
\begin{align}\label{eq:effectivedefocus}
    \Delta f_\mathrm{eff} \approx \Delta f + \frac{M\Delta z}{2},
\end{align}
where $M\Delta z$ is the thickness of the scattering volume $v$.

\subsubsection{The missing phase problem.}
\label{sec:missingphase}

Recall in Eq.~\eqref{eq:g} that the phases of the holograms are lost upon measurement.
It is non-trivial to recover the complex-valued exitwave, $\psi_\mathrm{exit}$, directly from the real-valued image, $g$, as Eq.~\eqref{eq:g} is not invertible.
This is commonly known as the \emph{missing phases problem} in electron microscopy \citep{Kirkland1984-ok}.

If the phases of the holograms at the detector, $\psi_\mathrm{det}$, can be retrieved, then the complex-valued exitwaves, $\psi_\mathrm{exit}$, needed for DOFS can be recovered by backpropagation:
\begin{align}\label{eq:backprop}
    \psi_\mathrm{exit}\qty(\vb{x}) = \psi_\mathrm{det}\qty(\vb{x}) \otimes^{-1} h\qty(\vb{x}).
\end{align}
This backpropagation in Eq.~\eqref{eq:backprop} essentially removes the microscopy aberrations from the images but is only possible if the phases of the holograms are retrieved.

While CTF correction algorithms in cryo-EM also remove the aberrations from particle images, these algorithms use an alternative approach to work around the missing phase problem.
Rather than solving Eq.~\eqref{eq:g} directly, a common strategy is to perform a first-order expansion and assume a proportionality between the real and imaginary parts of the projected potential, $v_z$ (see Supplementary Information).
This linearization produces the familiar CTF, which can be removed from the images via deconvolution algorithms \citep{Penczek2010-iw, singer2020computational, Nakane2020-lo, yip2020atomic}.
However, such CTF correction methods do not recover the phases of these complex-valued exitwaves, $\psi_\mathrm{exit}$.
Without these phases, DOFS cannot proceed and ESCC becomes non-trivial.
In the next section, we describe how to recover these exitwaves using computational phase retrieval.

\subsubsection{Phase retrieval.}

To retrieve the phases of a single exitwave, we first recast Eq.~\eqref{eq:g} into a minimization problem:
\begin{align}\label{eq:phasing}
    \min_{\psi_{\mathrm{exit}}} \norm{\sqrt{g} - \qty|\psi_\mathrm{exit} \otimes h|}^2_2 \equiv \min_{\psi_{\mathrm{exit}}} f\qty(\psi_{\mathrm{exit}}),
\end{align}
where $f$ is the squared L2-norm loss.
Taking the derivative of the objective in Eq.~\eqref{eq:phasing} with respect to the exitwave, we obtain the gradient:
\begin{align}\label{eq:2Dgrad}
    \pdv{f}{\psi_{\mathrm{exit}}} = -2W \qty\Bigg{\qty\bigg[\underbrace{\vphantom{\frac{\psi_\mathrm{exit} \otimes h}{\qty|\psi_\mathrm{exit} \otimes h|}} \qty(\sqrt{g} - \qty|\psi_\mathrm{exit} \otimes h|)}_{\text{error term}}\times \underbrace{\frac{\psi_\mathrm{exit} \otimes h}{\qty|\psi_\mathrm{exit} \otimes h|}}_{\text{phasor}}] \otimes^{-1} h}.
\end{align}
The gradient in Eq.~\eqref{eq:2Dgrad} has an intuitive interpretation: it rescales the current phases of the hologram (phasor term) by the error between the measured and estimated image (error term) and backpropagates this to the exitplane (deconvolution term).
Eq.~\eqref{eq:2Dgrad} is also known as the reweighted amplitude flow (RAF) algorithm, where the $W$ term is introduced to accelerate convergence \citep{Wang2018-cr} (details in Supplementary Information).

Multiple exitwaves (at $N$ different orientations of the particle) are needed to form a 3D volume.
Rather than performing $N$ independent 2D gradient descents for each measured image \citep{Gureyev2022-gt}, we constrain these exitwaves to emerge from the same 3D volume.
This constraint avoids the poor convergence observed in single-shot phase retrieval \citep{Gureyev2022-gt}.
To implement this constraint, we introduce the concept of a complex-valued surrogate 3D volume, $\psi_\mathrm{SV}\qty(x,y,z)$, that uses the PA to produce holograms whose intensities match those of measured images.
This surrogate volume is the tomographic reconstruction of the current estimates of the exitwaves.
However, it is important to note that this surrogate volume is not strictly the true 3D structure of the particle. 
Only when the particle is sufficiently small (i.e., the PA is valid), will the former volume approach the latter.
Similarly, the 2D gradients calculated for each particle using Eq.~\eqref{eq:2Dgrad} are averaged together to update the individual voxels of this surrogate volume.
This volumetric derivative, $\pdv{f}{\psi_\mathrm{SV}}$, can be used in the gradient descent update:
\begin{align}\label{eq:3Dgrad}
    \psi^{k+1}_\mathrm{SV} = \psi^k_\mathrm{SV}- \eta\pdv{f}{\psi^k_\mathrm{SV}},
\end{align}
where $\eta$ is a user-defined stepsize.

\subsection{Ghostbuster algorithm}

\begin{figure}[htbp]
    \centering
    \includegraphics[width=\textwidth]{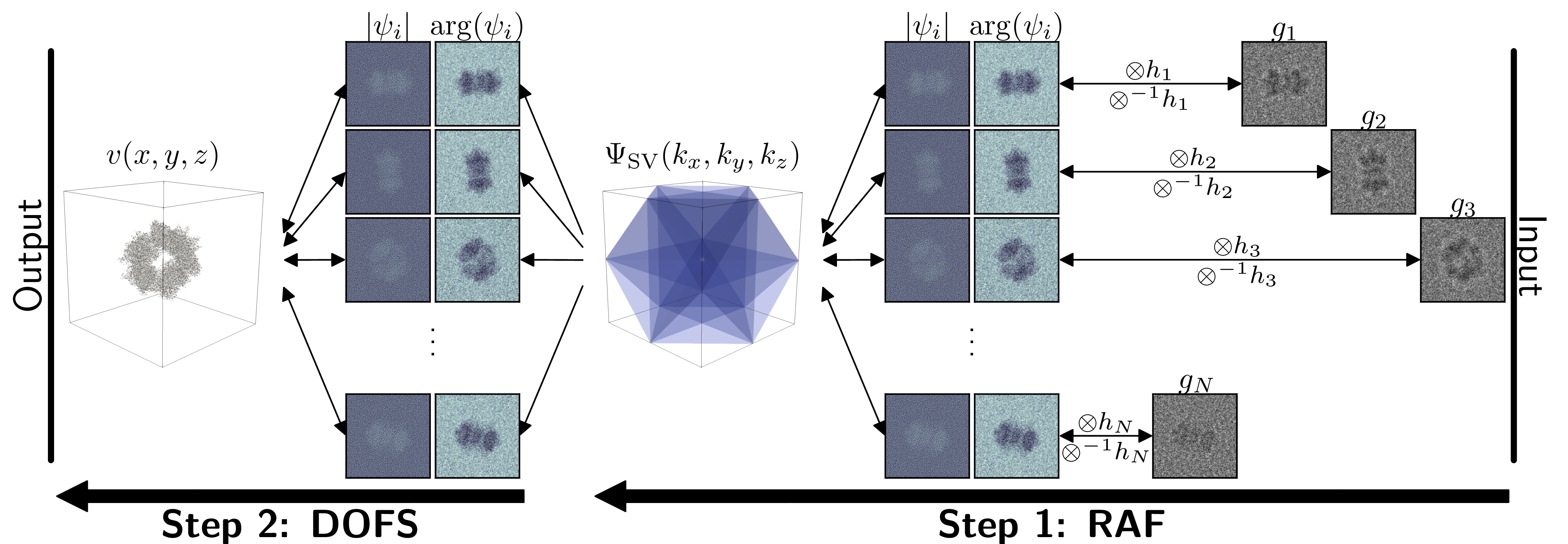}
    \caption{
    Schematic of the Ghostbuster algorithm. 
    In the first step, the RAF algorithm is used to recover the 2D complex-valued exitwaves, $\psi_i$ (omitting the `exit' subscript for brevity), from real-valued particle intensities, $g_i$.
    The operations $\otimes h_i$ and $\otimes^{-1}h_i$ represent the forward and backward propagation of the waves between the exitplane and the $i^{\text{th}}$ image plane. 
    At every RAF iteration, the current estimates of the 2D exitwaves concur into a surrogate volume, $\psi_\mathrm{SV}\qty(x,y,z)$, via the projection approximation (depicted here in the Fourier domain as $\Psi_\mathrm{SV}\qty(k_x,k_y,k_z)$ using the Fourier slice theorem).
    After completing all RAF iterations, the phase-retrieved exitwaves are then used in DOFS to recover the 3D particle, $v$, in the second step.}
    \label{fig:ghostbuster}    
\end{figure}

Our proposed Ghostbuster algorithm combines the RAF and DOFS algorithms described in previous sections to form an end-to-end pipeline that reconstructs a 3D particle from the measured, phaseless images (Fig.~\ref{fig:ghostbuster}).
The surrogate volume, $\psi_\mathrm{SV}\qty(x,y,z)$, is first recovered using phase retrieval via RAF.
The 2D exitwaves can be extracted from this exitwave volume via tomographic projection.
We then normalize these exitwaves such that the unscattered beam has unit amplitude and zero phase, which is numerically consistent with Eq.~ \eqref{eq:linear_multislice}.
These phase-retrieved exitwaves are then used by DOFS to stretch the reconstruction's DOF.
This stretching effectively eliminates the ghost-like artifacts seen in PA-based reconstructions, which we term ghostbusting.
Note that Ghostbuster assumes that the defocus and orientation of each particle are correctly inferred from cryo-EM software.

\section{Results and discussion}

In this section, we perform several numerical experiments to demonstrate the efficacy of both DOFS and Ghostbuster using multislice simulated data.
For the following simulations, we used 50 iterations for RAF and 5000 iterations for DOFS, with stepsizes of $\mu=\eta=1$.
RAF is initialized using the pseudoinverse algorithm based on CTF correction \citep{Penczek2010-iw}, whereas DOFS is initialized with zeros.

\subsection{Data generation}

Here, we simulate multislice datasets for both exitwaves as well as cryo-EM images using the Methanococcoides burtonii rubisco complex (PDB ID: 5MAC) \citep{pdb-5mac} and the biotin-bound streptavidin (PDB ID: 6J6J) \citep{Fan2019-uj}.
For a given particle, we first sample its 3D real-valued scattering potential, $\tilde{v}$, on a $M \times M \times M$ grid from its PDB file obtained from the Protein Data Bank \citep{Berman2000-rg} (see Table.~\ref{tab:sampling} for details).
For Sec.~\ref{sec:ice} and \ref{sec:endtoend}, we also add the scattering potential of random ice to simulate cryo-EM particles.

\begin{table}[]
    \centering
    \begin{tabular}{@{}rcccc@{}}
        \toprule
        \multicolumn{1}{l}{}                 & \multicolumn{2}{c}{Sec.~\ref{sec:size}} & Sec.~\ref{sec:ice}        & Sec.~\ref{sec:endtoend} \\ \midrule
        \multicolumn{1}{r|}{Particle}        & 5MAC     & \multicolumn{1}{c|}{6J6J}    & \multicolumn{1}{c|}{5MAC} & 5MAC                    \\
        \multicolumn{1}{r|}{No. slices, $M$} & 512      & \multicolumn{1}{c|}{256}     & \multicolumn{1}{c|}{256}  & 256                     \\
        \multicolumn{1}{r|}{Voxel size, $\Delta z$} & \SI{0.5}{\angstrom} & \multicolumn{1}{c|}{\SI{0.5}{\angstrom}} & \multicolumn{1}{c|}{\SI{1}{\angstrom}} & \SI{1}{\angstrom} \\ \bottomrule
    \end{tabular}
    \caption{Particle and sampling parameters for the generated datasets.}
    \label{tab:sampling}
\end{table}

To include the effects of absorption, we use an amplitude contrast ratio of $\alpha = 0.1$ \citep{Humphreys1968-da} to scale the real and imaginary parts of the complex-valued potential, $v$, using
\begin{align}\label{eq:alpha}
    v\qty(x,y,z) = \sqrt{1-\alpha^2}\tilde{v}\qty(x,y,z) + \iu \alpha \tilde{v}\qty(x,y,z).
\end{align}
This complex-valued potential is further scaled with an interaction parameter of $\sigma_e=$\SI{7.29e-4}{\radian \angstrom^{-1}\V^{-1}} \citep{Kirkland2010-kj}.
A \SI{200}{\keV} plane wave electron beam is incident on 3500 different orientations of this volume and the multislice method is used to generate the exitwaves.

To simulate the 3500 cryo-EM images for Sec.~\ref{sec:endtoend}, the exitwaves above are further aberrated by a spherical aberration of $C_s = $ \SI{2}{\mm} and defocus parameters ranging between \SI{0}{\um} to \SI{3}{\um}: for each of the seven average defocuses ($\Delta f_\mathrm{avg}=\qty{0, 0.5, \ldots, 3}$ \SI{}{\um}), we added random jitter sampled from a uniform distribution of $\mathrm{U}[-12.8, 12.8]$ \SI{}{\nm} (to simulate ice thickness of \SI{25.6}{\nm}) to generate 500 particles for each average defocus.
The spatial frequencies of these holograms are further dampened using a Gaussian envelope with $B$-factor of \SI{64}{\angstrom^{2}} which empirically models effects such as partial coherence, particle damage, movement, misalignment, etc.
The intensities of these resultant holograms are scaled by an exit electron dose of \SI{20}{e/\angstrom^{2}} before adding Poisson noise to create the particle stack for 3D reconstruction.

\subsection{Effect of particle size on DOFS}
\label{sec:size}

\begin{figure}[htbp]
    \centering
    \includegraphics[width=\textwidth]{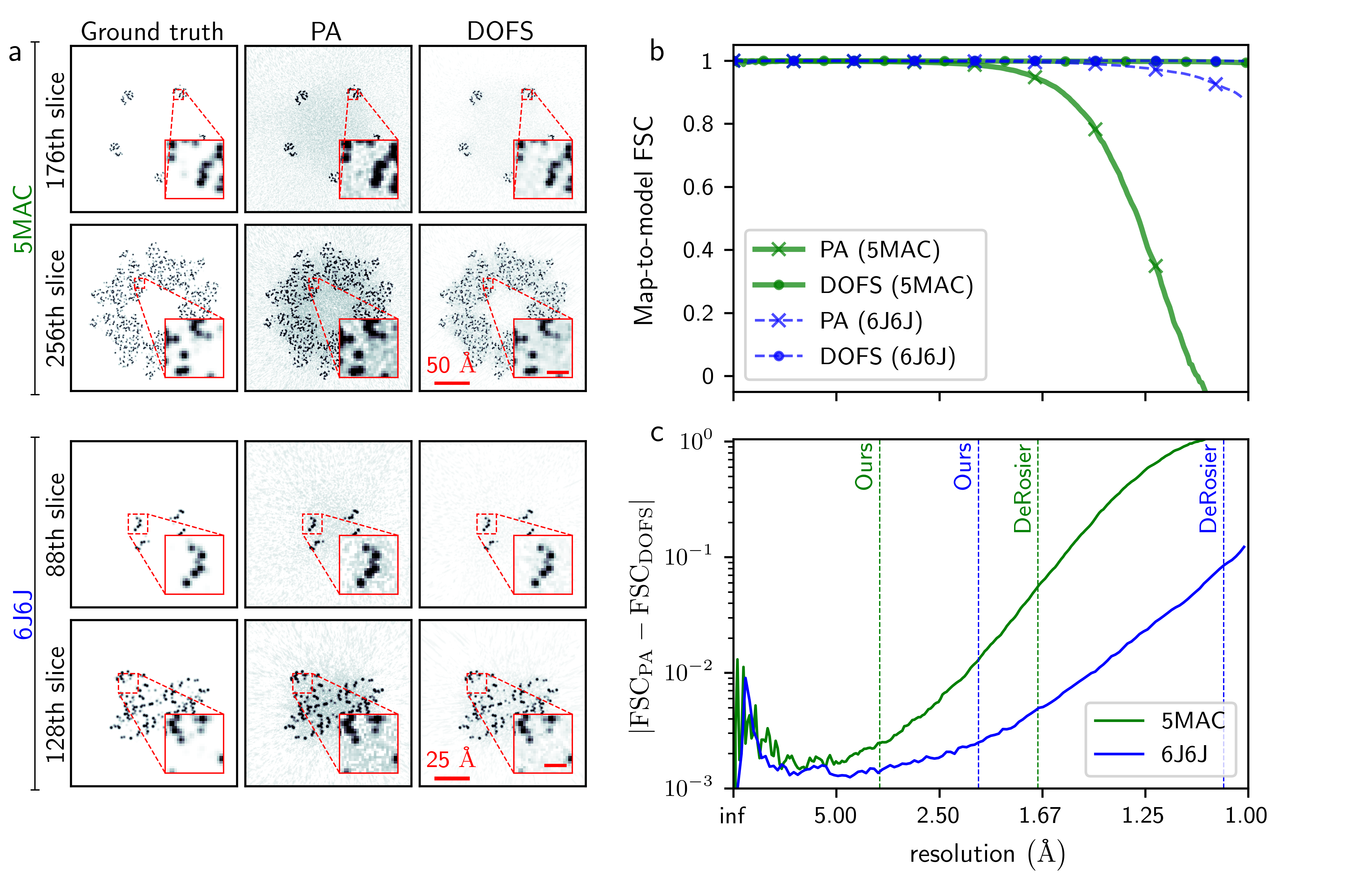}
    \caption{Reconstruction of 5MAC (\SI{160}{\angstrom} in diameter) and 6J6J (\SI{60}{\angstrom} in diameter) with ground truth exitwaves without ice.
    (a) Slices of the reconstructions from PA and DOFS compared against the ground truth (model).
    The scale bars within the insets correspond to \SI{5}{\angstrom}.
    (b) Map-to-model FSCs of the reconstructions from PA ($\times$) and DOFS ($\bullet$) for both 5MAC (green) and 6J6J (blue).
    (c) The absolute difference between the PA and DOFS FSCs for both 5MAC and 6J6J.
    DeRosier's rule-of-thumb estimates the PA to remain valid up to a resolution of \SI{1.69}{\angstrom} and \SI{1.05}{\angstrom} for 5MAC and 6J6J respectively.
    In contrast, our ghost-based criterion at 0.5-intensity-ratio estimates this limit to be \SI{3.52}{\angstrom} and \SI{2.10}{\angstrom} for 5MAC and 6J6J respectively.
    These reconstructions and their FSCs computed here are done without masks.}
    \label{fig:compare_size}
\end{figure}

Recall in Fig.~\ref{fig:resolution} that the resolution attainable using the PA decreases for thicker particles.
Here, we compare the 3D reconstructions between PA and DOFS from the groundtruth complex-valued exitwaves without ice for both 5MAC and 6J6J (Fig.~\ref{fig:compare_size}).
These particles have an approximate diameter of \SI{160}{\angstrom} and \SI{60}{\angstrom} respectively.

As expected, the appearance of ghosts in the PA is more apparent in the thicker particle (5MAC) compared to the thinner particle (6J6J).
The map-to-model Fourier shell correlation (FSC) curves reflect this observation as a loss in resolution due to the thickness of the particle.
In this highly idealized 5MAC example, it is clear that a \SI{1}{\angstrom} resolution reconstruction is impossible with the PA.
In both cases, the DOFS FSC curves remain close to one at all spatial resolutions up to Nyquist.
Visually, these DOFS reconstructions are able to bust most of the ghosts away, where the residual ghosts arise from the first Born approximation assumed in DOFS.

Fig.~\ref{fig:compare_size}c also includes the resolution estimation using both DeRosier's rule-of-thumb \citep{DeRosier2000-pd} and our ghost-based criterion described in Sec.~\ref{sec:dof}.
From the logarithmic deviation of the FSCs between DOFS and PA, we see that our criterion more precisely captures the spatial resolution beyond which the PA breaks down.

\subsection{Effects of ice on DOFS}
\label{sec:ice}

\begin{figure}[htbp]
    \centering
    \includegraphics[width=\textwidth]{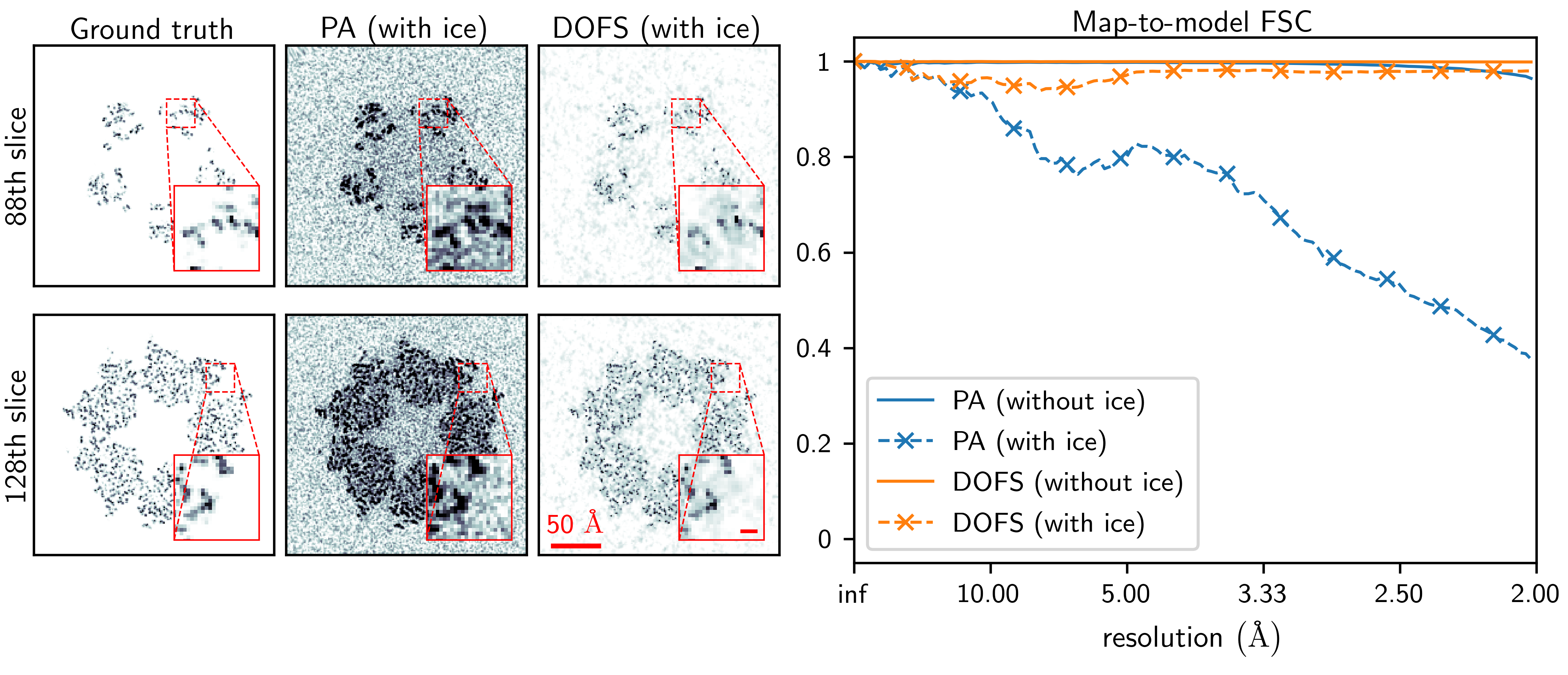}
    \caption{Reconstruction of 5MAC with (slices depicted above) and without ice (see Figure \ref{fig:compare_size}) given the ground truth complex-valued exitwaves with phases, which cannot be directly measured in experiments.
    The scale bar within the inset corresponds to \SI{5}{\angstrom}.
    We did not apply any masks to these reconstructions and their computed FSCs.}
    \label{fig:compare_ice}
\end{figure}

Here, we investigate the effects of ice on the resolution of particle reconstructions.
The presence of ice introduces random phase errors in the exitwaves, which will naturally degrade these reconstructions.
Nonetheless, these random errors from ice will be averaged away when reconstructing from multiple exitwaves since each will contain different and uncorrelated ice.
This averaging is shown in Fig.~\ref{fig:compare_ice} where we reconstruct 5MAC using both PA and DOFS.

By ignoring the curvature of the Ewald sphere, the PA reconstruction becomes more susceptible to ice noise compared to DOFS (see Sec.~\ref{sec:equivalence} for equivalence to ESCC).
During reconstruction, any Fourier component of the particle's structure is interpolated from the nearest complex-valued Ewald sphere sections.
Since each Ewald sphere section contains random phase errors from a different set of ice molecules, with sufficiently many sections (i.e. views of the particle), these phase errors for any Fourier component should be ideally averaged away during interpolation.
The PA, however, flattens these Ewald sphere sections and hence interpolates/updates 
into the neighboring Fourier components instead of the correct ones.
Consequently, this results in additional interpolation errors that deteriorate its resolution compared to DOFS.

\subsection{Ghostbuster reconstructions in realistic cryo-EM conditions}
\label{sec:endtoend}

\begin{figure}[htbp]
    \centering
    \includegraphics[width=\textwidth]{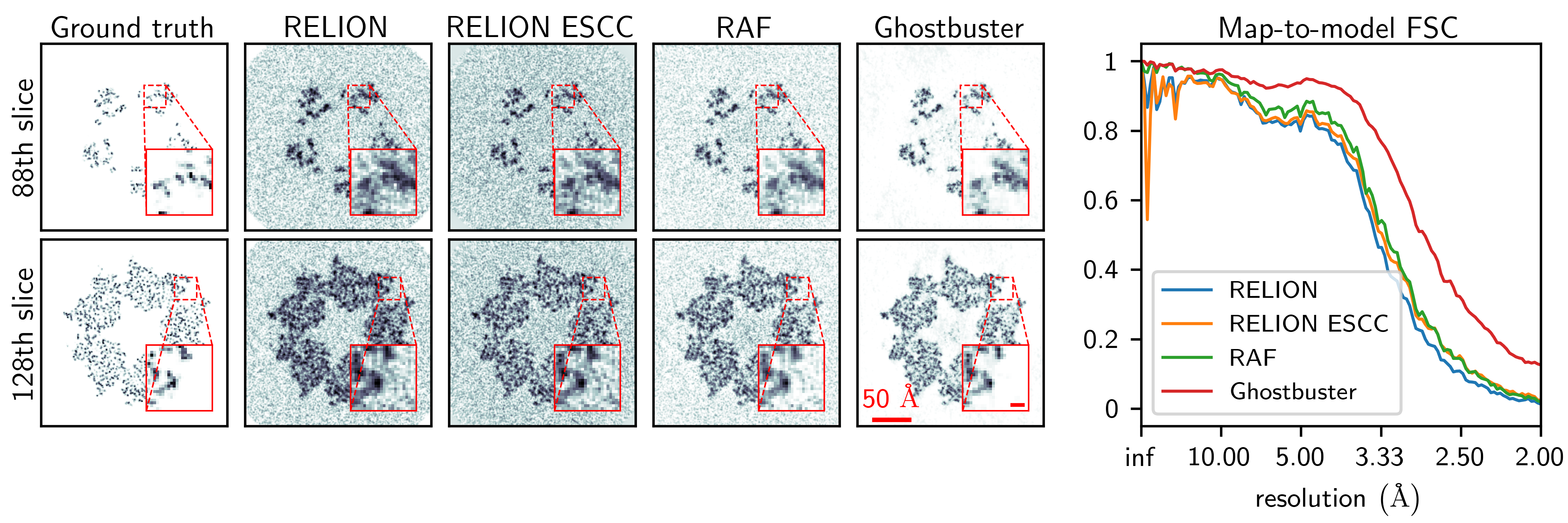}
    \caption{End-to-end Ghostbuster reconstruction of 5MAC from simulated cryo-EM images.
    The scale bar within the inset corresponds to \SI{5}{\angstrom}.
    All the reconstructions here have the same colormap except the ground truth (dynamic range is compressed).
    The reconstructions are not masked here.}
    \label{fig:recon}
\end{figure}

In this section, we test the Ghostbuster algorithm on a simulated cryo-EM particle stack (Fig.~\ref{fig:recon}).
We compare its results against the current state-of-the-art reconstruction software, RELION \citep{Kimanius2021-gd}, with and without ESCC.
RELION's reconstruction (without ESCC) essentially performs CTF correction on the particle stack followed by reconstruction using the PA.
Without accounting for the diffraction physics within the particle, RELION's reconstruction results in the ghostly artifacts seen in the slices in Fig.~\ref{fig:recon}.
These ghosts are slightly reduced when RELION employs ESCC (which inserts these CTF-corrected particles as curved surfaces in the Fourier volume).
This reduction in ghosts is correlated with the marginal increase in resolution in the map-to-model FSC curves.

Compared to RELION, Ghostbuster effectively eliminates these ghosts, as seen from the slices in Fig.~\ref{fig:recon}, and reports a better map-to-model FSC.
Ghostbuster's better reconstruction can be attributed to the complex-valued exitwaves recovered to stretch the DOF described in Sec.~\ref{sec:DOFS}.
Such stretching is not possible through CTF correction, which does not recover these complex-valued exitwaves (Sec.~\ref{sec:missingphase}).

To better see the effects of these ghosts, the map-to-model FSCs in Fig.~\ref{fig:recon} are computed without masks.
Ideally, even if we had the ground truth model as a mask, Ghostbuster's reconstruction still bests that of RELION's ESCC within this mask (see Supplementary Information).
In practice, users, however skillful they might be, may create poorer masks. 
Ghostbusting can reduce the dependence on such masking.

We speculate that RELION is ultimately shackled by CTF correction, which recall from Sec.~\ref{sec:missingphase}, solves a first-order approximation of the phase retrieval problem.
As a result, the FSC of RAF (phase retrieval with PA) is consistently above that of RELION's (without ESCC), even with masking (see Supplementary Information).
This suggests that computational phase retrieval sets the upper bound for CTF correction.

\section{Conclusion}
\label{sec:conclusion}

Ghostbuster is a diffraction tomography algorithm for 3D SPR.
Here we showed the appearance of ghostly artifacts in 3D reconstructions when scattering within the particle is ignored, even for previously assumed thin particles.
These ghosts are eliminated with Ghostbuster by first recovering the complex-valued exitwaves via phase retrieval, followed by DOFS which accounts for kinematic scattering to reconstruct the particle.
Reconstructions from multislice simulations of 5MAC particles demonstrated this ghostbusting, which improved resolution compared to RELION.

We also formulated a relationship between DOF and atomic separation in Sec.~\ref{sec:dof} based on the emergence of a ghost from two-atom interference.
This relationship can be interpreted as the PA's resolution limit for different particle thicknesses solely due to diffraction physics.
However, PA's resolution limit is overshadowed by the loss in resolution due to noise and uncertainties in experiments.
Hence, we use simulated data in this study to isolate and demonstrate PA's resolution limit.
Using the same data, we extend beyond PA's resolution limit by accounting for the diffraction physics within the particle (Sec.~\ref{sec:size}).

From our findings, we expect the ghosts to be more apparent for either thicker samples (e.g., in tomography \citep{Young2023-ft}) or cryo-EM at lower electron energies (e.g., 100 keV \citep{McMullan2023-da}). 
Both areas are being heavily developed to allow for in-situ structural understanding and for cheaper screening microscopes respectively \citep{Naydenova2019-ze}. 
The use of Ghostbusting can help complement both developments.

While ghosts are empirically observed in cryo-EM SPR, they tend to be suppressed by users' expert masking.
Here we show that Ghostbusting can minimize the need for such manual masking.
Admittedly, the improvements demonstrated in this paper may be lessened because of experimental uncertainties such as errors in the inferred defocuses and orientations of particles.
Nonetheless, we are optimistic that our Ghostbuster algorithm, which can be integrated seamlessly into existing cryo-EM imaging pipelines as a final refinement step without collecting additional data, can improve imaging outcomes.

From this paper, we see how phase retrieval can help usher in computational lensing into cryo-EM.

\paragraph{Acknowledgements} 
The authors would like to acknowledge the computational resources from NUS Centre for Bio-Imaging Sciences, and generous support from the NUS Institute of Mathematical Sciences.

\paragraph{Data and code availability}
Both datasets and the codes for Ghostbuster, DOFS, and RAF are available at \url{https://doi.org/10.5281/zenodo.10359522}.

\bibliographystyle{unsrtnat}


\newpage

\appendix

\begin{center}
    \textbf{{\huge Supplementary Information}}
\end{center}
\numberwithin{equation}{section}

\setcounter{equation}{0}

\section{Derivation of the asymptotes for the depth of field for a coherent system}
We assume that an atom located $z = 0$, centered at $x = y = 0$, can be described by a 2D Gaussian:
\begin{align}
    g\qty(x,y) = e^{-\frac{x^2 + y^2}{2 \sigma^2}},
\end{align}
where $\sigma$ is the standard deviation of the Gaussian which relates to the size of the atom.
Here, we use the definition of the size or resolution of this atom as the reciprocal of the 1/2 width of a Gaussian in Fourier space, i.e.:
\begin{align}
    \sigma = \frac{\mathrm{res}}{2\sqrt{2\log{2}}}.
\end{align}
Specifically, we choose $\mathrm{res} =$ \SI{0.7}{\angstrom} which corresponds to the atomic radius of carbon \citep{Slater1964-ed}.

The electron wave from this atom after propagating a certain distance $z$ is given by
\begin{align}\label{eq:convolve}
    g_z \qty(x,y) = g\qty(x,y) \otimes h\qty(z),
\end{align}
where $h$ is the Fresnel propagator, defined by its Fourier transform as
\begin{align}\label{eq:fresnelH}
    H\qty(k_x, k_y) = e^{\iu \pi z \lambda k^2},
\end{align}
where $\qty|k| = \sqrt{k_x^2+k_y^2}$.
Substituting Eq.~\eqref{eq:fresnelH} into Eq.~\eqref{eq:convolve} and using the convolution theorem, we obtain
\begin{align}
    g_z\qty(x,y) = \underbrace{\frac{\sigma^2}{\qty(\sigma^2 - \frac{i z \lambda}{2\pi})}}_{\mathrm{scalar}}\underbrace{\exp(-\frac{x^2 + y^2}{2\sigma^2 + \frac{z^2 \lambda^2}{2\pi^2\sigma^2}})}_{\mathrm{amplitude}} \underbrace{\exp(-\frac{\qty(x^2 + y^2)\frac{i z \lambda}{2\pi \sigma^2}}{2\sigma^2 + \frac{z^2 \lambda^2}{2\pi^2\sigma^2}})}_{\mathrm{phase}}.
\end{align}

To define the depth of field for a coherent system, we place two such Gaussian atoms at $x = d$ and $x = -d$, i.e. a distance of $2d$ apart from each other.
The resultant propagated electron wave, $g_z^\mathrm{atoms}$, is therefore
\begin{align}
    g_z^\mathrm{atoms}\qty(x,y) =&\, g_z\qty(x-d,y) + g_z\qty(x+d,y) \nonumber \\
    =&\, \underbrace{\frac{2\sigma^2\cosh\qty(\frac{dx + \frac{idx z \lambda}{2\pi\sigma^2}}{\sigma^2 + \frac{z^2 \lambda^2}{4\pi^2\sigma^2}})}{\qty(\sigma^2 - \frac{i z \lambda}{2\pi})}}_{\text{scalar}} \underbrace{\exp(-\frac{x^2+ d^2 + y^2}{2\sigma^2 + \frac{z^2 \lambda^2}{2\pi^2\sigma^2}})}_{\text{amplitude}} \underbrace{\exp(-\frac{\qty(x^2+ d^2 + y^2)\frac{i z \lambda}{2\pi \sigma^2}}{2\sigma^2 + \frac{z^2 \lambda^2}{2\pi^2\sigma^2}})}_{\text{phase}},
\end{align}
and its corresponding intensity is
\begin{align}
    I_z^\mathrm{atoms}\qty(x,y) &= \qty|g_z^\mathrm{atoms}\qty(x,y)|^2 \nonumber\\
    &= \qty|\frac{2\sigma^2\cosh\qty(\frac{dx + \frac{idx z \lambda}{2\pi\sigma^2}}{\sigma^2 + \frac{z^2 \lambda^2}{4\pi^2\sigma^2}})}{\qty(\sigma^2 - \frac{i z \lambda}{2\pi})}|^2 \exp(-\frac{x^2+ d^2 + y^2}{\sigma^2 + \frac{z^2 \lambda^2}{4\pi^2\sigma^2}}).
\end{align}

Due to interference effects, a central ghost atom arises on the $z$-axis (i.e., $x = y = 0$).
At a given $z$ propagated distance, we are interested in the ratio, $R$, of the central ghost intensity with respect to that of the original Gaussians at $x=\pm d, y=0$:
\begin{align}\label{eq:R}
    R\qty(z) &= \frac{I_z^\mathrm{atoms}\qty(x=0,y=0)}{I_z^\mathrm{atoms}\qty(x=\pm d,y=0)} \nonumber\\
    &= \frac{2\exp(\frac{d^2\sigma^2}{u\qty(z)})}{\cosh \frac{2\sigma^2d^2}{u\qty(z)} + \cos \frac{d^2 z \lambda}{u\qty(z)\pi}}
\end{align}
where we have defined $u\qty(z) \equiv \sigma^4 + \frac{z^2 \lambda^2}{4\pi^2}$.

To derive the equation of the asymptotes, we are interested in the behavior of $R$ when $z\rightarrow\infty$.
When $z$ is large, the constant in $u\qty(z)$ may be neglected, giving us
\begin{align}
    u\qty(z) &\approx \frac{z^2\lambda^2}{4\pi^2}.
\end{align}
Substituting this into Eq.~\eqref{eq:R}, we have
\begin{align}
    R &\approx \frac{2\exp(\frac{4\pi^2d^2\sigma^2}{z^2\lambda^2})}{\cosh \frac{8\pi^2\sigma^2d^2}{z^2\lambda^2} + \cos \frac{4\pi d^2 \lambda}{z\lambda^2}}.
\end{align}

Observing the linear trend of the ghost-based criteria, we assume that at large values of $z$, the ratio $\frac{d}{z}\rightarrow C$ where $C$ is some constant. This gives us
\begin{align}\label{eq:Rapprox}
    R &\approx \frac{2\exp(\frac{4\pi^2C^2\sigma^2}{\lambda^2})}{\cosh \frac{8\pi^2\sigma^2 C^2}{\lambda^2} + \cos \frac{4\pi C d \lambda}{\lambda^2}}.
\end{align}
Assuming that the argument in the $\cosh$ term is large, then the $\cosh$ term dominates in the denominator since $-1 \leq \cos\qty(\ldots)\leq 1$.
In addition, by definition of $\cosh w = \frac{\exp(w) + \exp(-w)}{2}$, when $w \gg 1$, then $\cosh w \approx \frac{\exp(w)}{2}$.
These simplify Eq.~\eqref{eq:Rapprox} to
\begin{align}
    \frac{1}{2}R \exp(\frac{8\pi^2\sigma^2 C^2}{\lambda^2}) &\approx 2\exp({\frac{4\pi^2C^2\sigma^2}{\lambda^2}})
\end{align}
Solving for $C$, we have
\begin{align}
    C &= \sqrt{\frac{\lambda^2}{4\pi^2\sigma^2}\log{\frac{4}{R}}}.
\end{align}

\begin{figure}[htbp]
    \centering
    \includegraphics[width=1\textwidth]{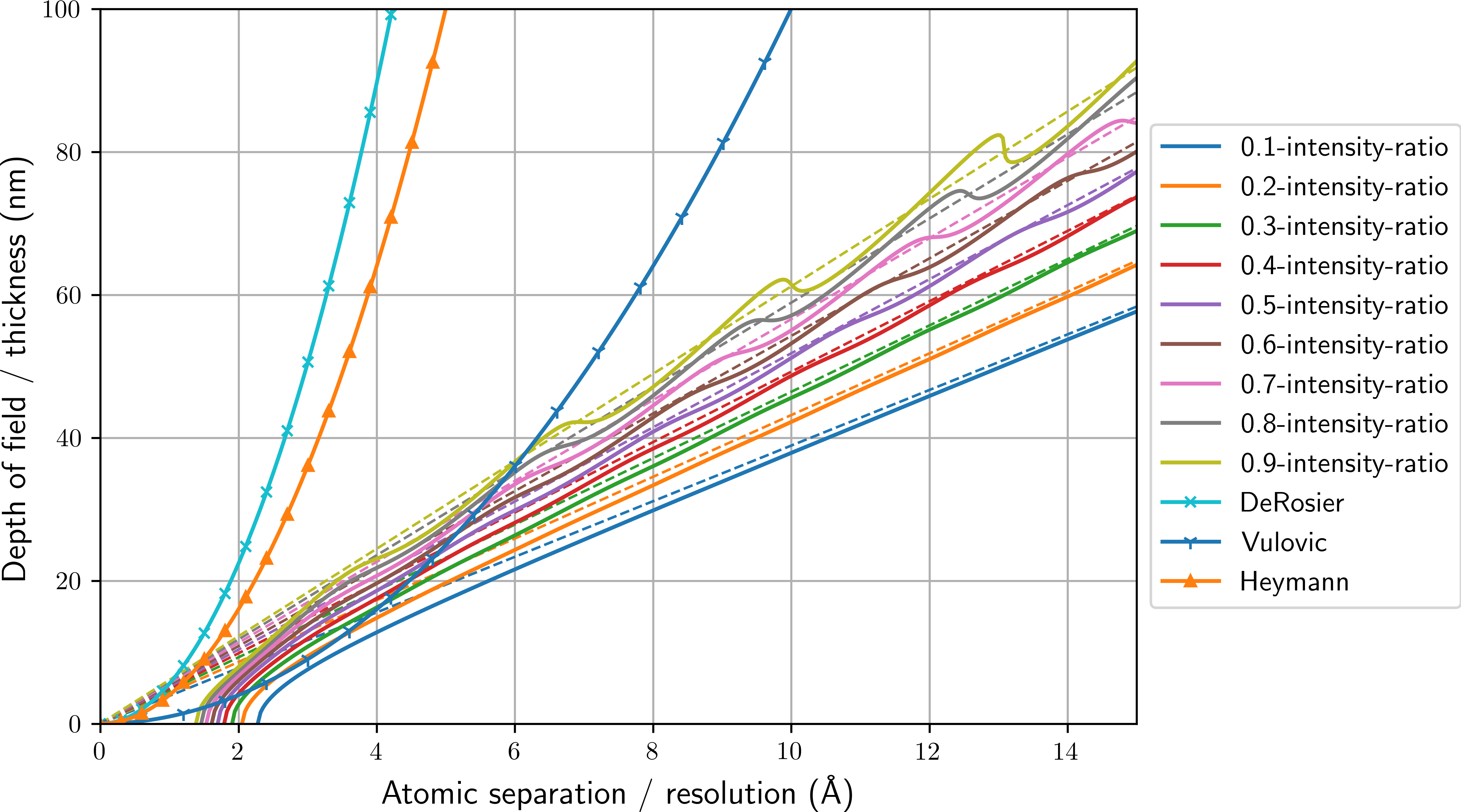}
    \caption{The DOF (or particle thickness) as a function of atom separation (or resolution).}
    \label{fig:allresolution}
\end{figure}

Therefore, the equations of the asymptotes are the straight lines:
\begin{align}
    z\qty(s) = \frac{d}{C} = \frac{s}{2C} = \frac{\pi\sigma }{\lambda \sqrt{\log{\frac{4}{R}} }} s, \qq{or} t\qty(s) = \frac{\pi\sigma }{\lambda \sqrt{\log{\frac{4}{R}} }} r,
\end{align}
where $s\equiv2d$ is the separation between the Gaussians at $z=0$, and $t\equiv z$ and $r \equiv s$ are the thickness and resolution particle respectively as commonly related via resolution estimates in \citep{DeRosier2000-pd, Vulovic2014-nd, Heymann2023-jp}.
We plot the ghost-based criteria for different $R$ values in Fig.~\ref{fig:allresolution}.

\section{Frozen phonon approximation for ghost formation}
Here we consider the effects of vibrating atoms on the formation of the central ghost atoms.

\begin{figure}[htbp]
    \centering
    \includegraphics[width=\textwidth]{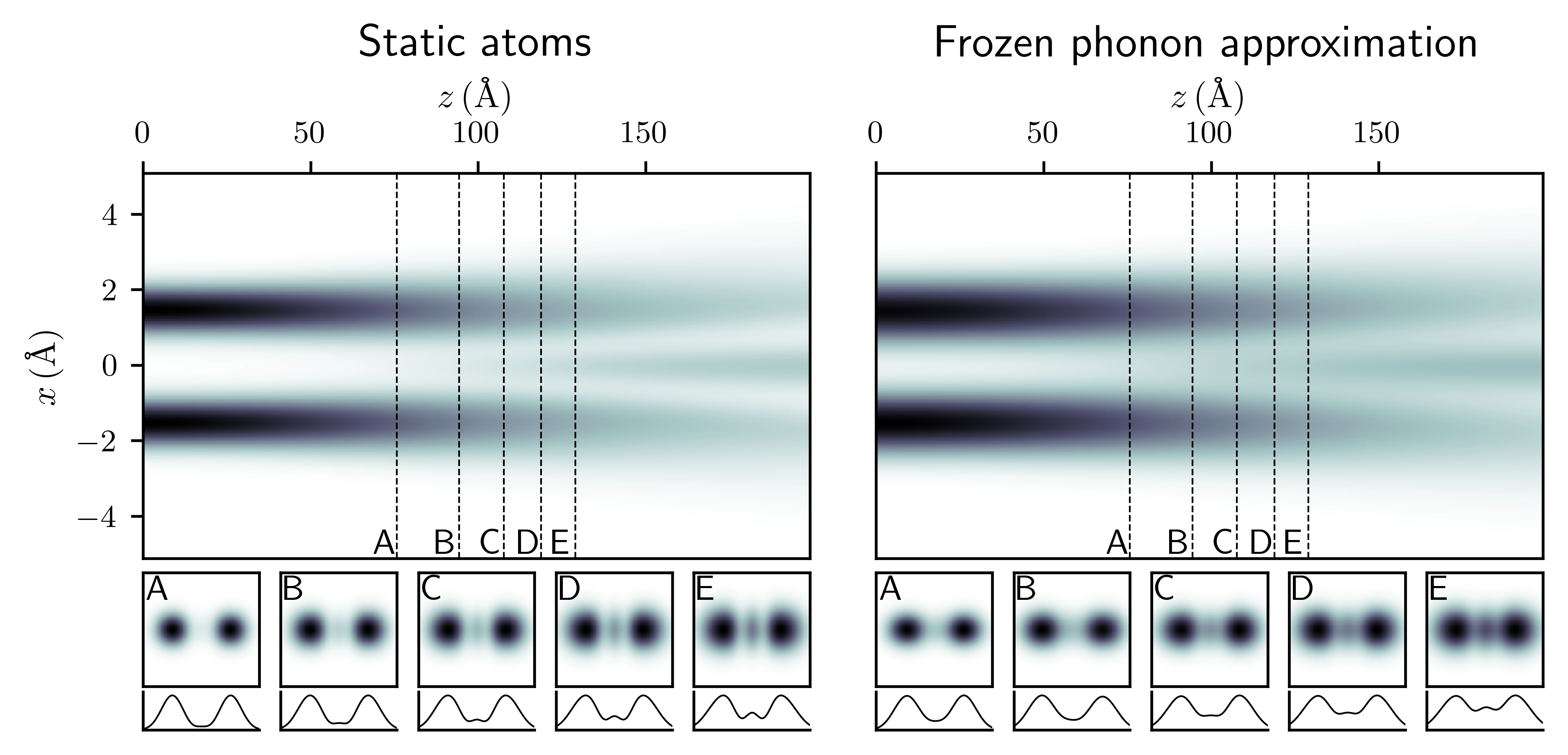}
    \caption{Formation of ghost atom for (left) static atoms and (right) with the frozen phonon approximation.}
    \label{fig:resolution_compare}
\end{figure}

Fig.~\ref{fig:resolution_compare} shows the appearance of the central ghost atom for both static atoms as well as under the frozen phonon approximation (FPA).
The FPA scenario is simulated by including random translations of the atoms sampled from a Gaussian distribution with a standard deviation of \SI{0.35}{\angstrom} (half the atomic radius of carbon).
We simulated the wave propagation of 100 trials and incoherently summed the waves together to obtain the FPA image.
This demonstrates that with atomic vibrations, ghosts become more prominent and effectively reduce the DOF further at a given resolution.

\section{Derivation of CTF correction equation}
The linear dependence between the real and imaginary parts of the complex-valued scattering potential, $v$, is a standard assumption in electron diffraction theory~\citep{Humphreys1968-da} and electron microscopy~\citep{Wade1992-bs,Wang2010-vv}.
This is often written as
\begin{align}\label{eq:potential}
    v\qty(x,y,z) = \sqrt{1-\alpha^2} \tilde{v}\qty(x,y,z) + \iu\alpha \tilde{v}\qty(x,y,z)
\end{align}
where $\tilde{v}$ is the real-valued scattering potential (commonly known as the `map'), and $\alpha$ is the amplitude contrast ratio.
For a wide range of materials, $\alpha$ is typically of the order of $\sim0.1$~\citep{Humphreys1968-da}.

The derivation of the CTF-correction equation begins by invoking the projection approximation, which projects the 3D complex scattering potential of the sample at a particular orientation along the $z$-axis:
\begin{align}\label{eq:projection}
    v_z\qty(\vb{x}) = \int v\qty(x,y,z) \dd{z} = \sqrt{1-\alpha^2} \tilde{v}_z\qty(\vb{x}) + \iu\alpha \tilde{v}_z\qty(\vb{x}),
\end{align}
where the $z$ subscripts denote an integral along the $z$-axis.
Under this assumption, the real and imaginary parts of the projected potential, $v_z$, retain the same proportionality described in Eq.~\eqref{eq:potential}.

The effects of the sample on the incident beam are then approximated as a transmission function of the form \citep{Vulovic2014-nd}:
\begin{align}\label{eq:t}
    \psi_\mathrm{exit}\qty(\vb{x}) \approx \exp[\iu\sigma_e v_z\qty(\vb{x})].
\end{align}
In addition, when $\qty|\sigma_e v_z| \ll 1$, the weak-phase approximation linearizes Eq.~\eqref{eq:t} to \citep{Vulovic2014-nd}
\begin{align}\label{eq:weakphase}
    \psi_\mathrm{exit}\qty(\vb{x}) \approx 1 + \iu\sigma_e v_z\qty(\vb{x}) + \order{v_z^2}.
\end{align}

Recalling that the image measured on the detector is $g = \qty|\psi_\mathrm{exit} \otimes h|^2$, substituting Eq.~\eqref{eq:weakphase} and expanding the modulus-square gives
\begin{align}\label{eq:g_linear}
    g\qty(\vb{x}) \approx 1 + \iu\sigma_e v_z\qty(\vb{x}) \otimes h\qty(\vb{x}) - \sigma_e v_z^*\qty(\vb{x}) \otimes h^*\qty(\vb{x}).
\end{align}
Note that higher-order terms of $\order{v_z^2}$ have been dropped in Eq.~\eqref{eq:g_linear}.

Using Eq.~\eqref{eq:potential} on the last result, the Fourier transform of $g$ becomes
\begin{align}\label{eq:G_linear}
    G\qty(\vb{k}) = \delta\qty(\vb{k}) + \sigma_e \tilde{V}_z\qty(\vb{k})\qty{\iu\sqrt{1-\alpha^2}  \qty\Big[H\qty(\vb{k}) - H^*\qty(\vb{-k})] - \alpha \qty\Big[H\qty(\vb{k}) + H^*\qty(\vb{-k})]},
\end{align}
where the capital letters are the Fourier transforms of the respective quantities. $H$, in particular, is known as the transfer function of the electron microscope, and $\vb{k}$ denotes the 2D spatial frequency coordinates.

The transfer function in electron microscopy is typically defined as \citep{Kirkland2010-kj}
\begin{align}\label{eq:transfer}
    H\qty(\vb{k}) = \exp[-\iu\chi\qty(\vb{k})],
\end{align}
where $\chi$ is the aberration function of the microscope and is assumed to be a real function. Many cryoEM software packages assume a model for $\chi$ and incorporate fitting algorithms to fit the parameters for the aberration model from the measured $g_n$. For the scope of this paper, we shall assume the simplest model for $\chi$ given as \citep{Kirkland2010-kj}
\begin{align}\label{eq:aberration}
    \chi\qty(\vb{k}) = 2\pi\qty(-\frac{1}{2}\Delta f \lambda k^2 + \frac{1}{4}C_s \lambda^3 k^4),
\end{align}
where $k=\qty|\vb{k}|$, $\lambda$ is the wavelength of the incident electron beam, and $C_s$ is the spherical aberration constant.

Substituting the transfer function, $H$, into Eq.~\eqref{eq:G_linear}, we arrive at the CTF-correction equation:
\begin{align}\label{eq:ctf}
    G\qty(\vb{k}) = 2\sigma_e \tilde{V}_z\qty(\vb{k}) \mathrm{CTF}\qty(\vb{k}),
\end{align}
where the CTF is defined as
\begin{align}
    \mathrm{CTF}\qty(\vb{k}) \equiv \sqrt{1 - \alpha^2} \sin \chi\qty(\vb{k}) - \alpha \cos \chi\qty(\vb{k}).
\end{align}
The $\delta$ term has been intentionally dropped in Eq.~\eqref{eq:ctf} as it represents the unscattered electron beam and is inconsequential in CTF correction algorithms.
Note that the assumed proportionality between the real and imaginary part of $v_z$ described in Eq.~\eqref{eq:projection} is ingrained in the CTF formulation.
This makes it non-trivial to decouple the projection approximation from the CTF formulation, which limits the ability of CTF correction methods to remove microscopy aberrations for thick samples even with Ewald sphere curvature correction strategies.

\section{Reweighted amplitude flow (RAF)}
The full details of the RAF algorithm may be found in \citep{Wang2018-cr}.
Here, we simply highlight some key points for completeness.
The RAF gradient for cryo-EM is given as
\begin{align}
    \pdv{f}{\psi_{\mathrm{exit}}} = -2W \qty\Bigg{\qty\bigg[\underbrace{\vphantom{\frac{\psi_\mathrm{exit} \otimes h}{\qty|\psi_\mathrm{exit} \otimes h|}} \qty(\sqrt{g} - \qty|\psi_\mathrm{exit} \otimes h|)}_{\text{error term}}\times \underbrace{\frac{\psi_\mathrm{exit} \otimes h}{\qty|\psi_\mathrm{exit} \otimes h|}}_{\text{phasor}}] \otimes^{-1} h}.
\end{align}
$W$ weighs the descent directions of the gradient for each datum to avoid misleading search directions due to poorly estimated signs for $\frac{\psi_\mathrm{exit} \otimes h}{\qty|\psi_\mathrm{exit} \otimes h|}$.
The ratio of $\frac{\qty|\psi_\mathrm{exit} \otimes h|}{\sqrt{g}}$ can be viewed as a confidence score for the corresponding gradient.
Hence, $W$ assigns larger weights to reliable gradients and smaller weights to erroneous ones through its definition:
\begin{align}
    W = \frac{1}{1 + \beta \frac{\sqrt{g}}{\qty|\psi_\mathrm{exit} \otimes h|}}.
\end{align}
In our implementation, we used $\beta = 0.8$.

\section{Masked reconstructions in realistic cryo-EM conditions}

\begin{figure}[htbp]
    \centering
    \includegraphics[width=\textwidth]{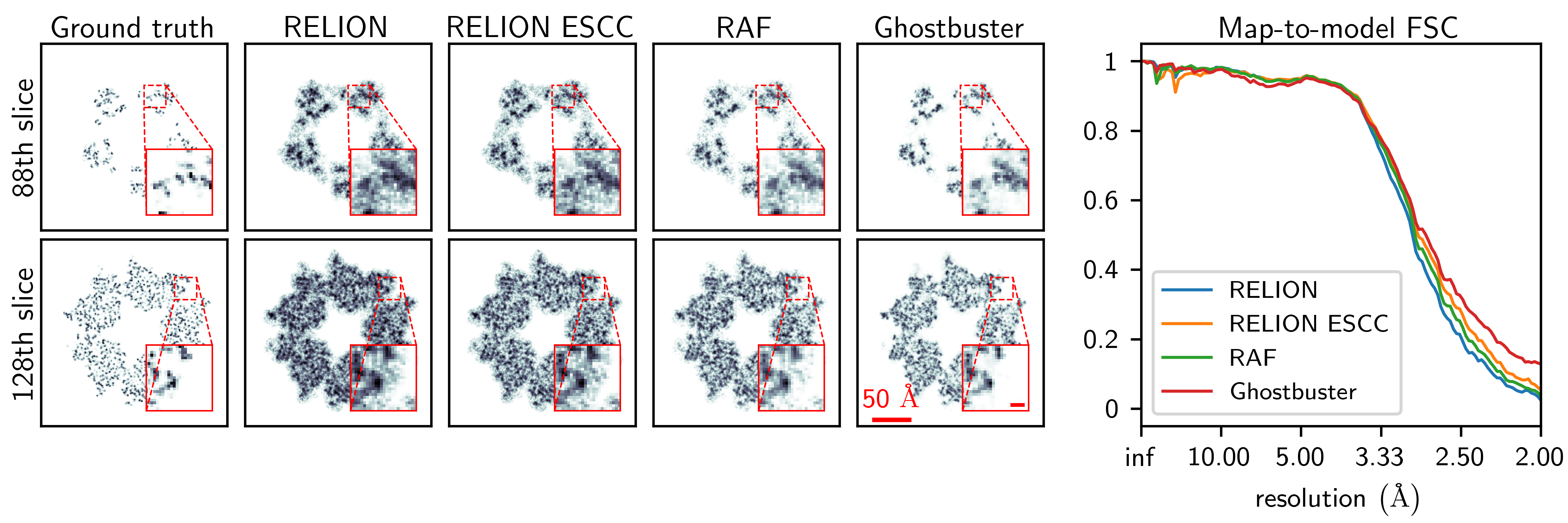}
    \caption{Reconstructions of 5MAC from simulated cryo-EM particles with masking.
    The mask was created using the ground truth map with additional three pixel dilation and three pixel cosine smoothness in RELION.}
    \label{fig:recon_masked}
\end{figure}

Even with a mask applied to the reconstructions, we still observe residual ghosts in the slices from RELION and RELION ESCC in Fig.~\ref{fig:recon_masked}.
Although the mask elevates the FSC curves for RELION reconstructions, it is evident that Ghostbuster remains as the reconstruction with highest fidelity from both the slices and FSCs.

\end{document}